\documentstyle[aps,multicol,epsfig,prl]{revtex}
\input epsf
\def\prb{Phys. Rev. B}
\def\prl{Phys. Rev. Lett.}

\def\be{\begin{equation}}
\def\ee{\end{equation}}
\def\ba{\begin{eqnarray}}
\def\ea{\end{eqnarray}}

\def\BSCCO{Bi$_2$Sr$_2$CaCu$_2$O$_{8+\delta}$}
\def\C60{A$_x$C$_{60}$}

\def\HgCu3{HgCa$_2$Cu$_3$O$_{8+y}$}
\def\HgCu4{HgBa$_2$Ca$_3$Cu$_4$O$_{10+y}$}
\def\TlCu3{Tl$_2$Ba$_2$Ca$_2$Cu$_3$O$_{10+y}$}
\def\TlCu4{Tl$_2$Ba$_2$Ca$_3$Cu$_4$O$_{12+y}$}
\def\TlCun{Tl$_2$Ba$_2$Ca$_{n-1}$Cu$_n$O$_{2n+4+y}$}
\def\HgCun{HgBa$_2$Ca$_{n-1}$Cu$_n$O$_{2n+2+y}$}

\def\BiCu3{Bi$_2$Sr$_2$Ca$_{2}$Cu$_3$O$_y$}

\begin{document}


\title{ Making High T$_c$ Higher:  A Theoretical Proposal}

\author{S.~A.~Kivelson}

\address
{  Deptment of Physics,
University of California,
Los Angeles, CA  90095}
\date{\today}

\maketitle

\begin{abstract}
\

There is considerable evidence that the highest $T_c$ obtainable in a
copper-oxide plane is limitted by the competition between two effects:
On the one hand, as the concentration of doped-holes, $ x$, is increased,
the pairing scale, which is related to the properties of a doped
Mott insulator, decreases.  On the otherhand, the superfluid density,
which controls the stiffness of the system to phase fluctuations,
vanishes as $x \to 0$, and increases with increasing $x$.  Optimal $T_c$
is obtained at a crossover from a phase ordering dominated regime at
small $x$ to a pairing dominated regime at large $x$.  If this
description is valid, then higher $T_c$'s can be obtained in an array of
coupled planes with different doped hole concentrations, such that a
high pairing scale is derived from the underdoped planes and a large
phase stiffness from the optimally or overdoped ones.

\
\end{abstract}
\begin{multicols}{2}

\narrowtext

This paper was prepared for a volume honoring the contributions of Zach
Fisk to materials physics on the occasion of his 60th birthday.  At the
associated ``Future of Materials Physics Workshop'' 
there were many quips (none by Fisk) concerning the question
of whether theoretical studies really are an essential part of this
future.  This paper is, in part, a response to those quips.  If we
believe that we have progressed in our theoretical understanding of the
physics of high temperature superconductivity, then we should be able to
use that understanding to provide guidance in the search for new, and
possibly higher temperature superconductors.  Given our incomplete
understanding of the physics of highly correlated electronic systems, it
is unreasonable to expect theory to point to specific materials, or
make quantitative predictions of superconducting transition temperatures.
However, I feel that there are aspects of the
theory of high temperature superconductivity that are sufficiently crisp
that they can provide qualitative information - what sort of materials
are likely to be high temperature superconductors, and what sorts of
modifications of the existing materials might lead to higher transition
temperatures.  In the final section of this paper, I make a specific 
proposal for raising the transition temperature of the cuprate superconductors which is
based on the theoretical notion that $T_c$ in the underdoped materials
is substantially supressed below its mean-field value by phase fluctuations.
At the very
least, the success or failure of such suggestions provides a non-trivial test of the
usefullness of some theoretical ideas.

\section{Background}

Superconductivity in metals is the result of two distinct quantum
phenomena, pairing and long-range phase coherence. In conventional
homogeneous superconductors the phase stiffness is so great that these
two phenomena occur simultaneously. On the other hand, in granular
superconductors and Josephson junction arrays, pairing occurs at the
bulk  transition temperature of the constituent metal, while long-range
phase coherence occurs, if at all, at a much lower temperature
characteristic of the Josephson coupling between superconducting
grains.   High temperature superconductivity\cite{BM} is hard to
achieve, even in theory, because it  requires that both scales be
elevated simultaneously--yet they are  usually incompatible.  Consider,
for example, the strong-coupling limit of the negative $U$ Hubbard
 or Holstein models.
Pairs have a large binding energy but, typically, they condense at a
very low temperature because of the large effective mass  of a tightly
bound pair - the effective mass is proportional to $|U|$ in the Hubbard
model and is exponentially large in the Holstein model.   (This also makes
the system susceptible to other types of
order, such as charge-density wave order, which compete with superconductivity.)

It was pointed out some time ago by Vic Emery and me\cite{nature} that the small
superfluid densities in the cuprate high temperature superconductors
implies that the competition between pairing and condensation plays a
significant role in the physics of these materials\cite{others}.  The superfluid
density in these materials is low, in part, because they are ``doped
Mott insulators'' in the sense that the superfluid density is roughly
proportional to the density of ``doped holes,'' $x$, rather than to the
full density of holes, $1+x$.

To make these considerations explicit,
imagine we start from a knowledge of the ground-state
properties of a superconductor and ask the question, ``At what
temperature would a given class of thermal fluctuations destroy the
superconducting long-range order?"  If the superfluid denisity is 
large, so condensation and pairing occur roughly simultaneously, as in 
the BCS theory, then
$T_{c}\approx T_{pair}$ where $T_{pair}$ is the temperature at which 
most of the pairs fall appart;   
$T_{pair}\equiv\Delta_0/2$ is proportional to a typical value
of the  zero temperature  superconducting gap,  $\Delta_0$.  
If, on the other hand, the superfluid density is small,
the effects of fluctuations of the phase of the superconducting order
determine $T_{c}\sim T_{\theta}$ where, at least in the case of 
layered superconductors, the
phase ordering temperature, $T_{\theta}\equiv \hbar^2 n_s/2m^*$,
can be deduced directly from the superfluid stiffness per layer,
$\hbar^2 n_sa_{c}/m^*$ and the mean spacing between layers, $a_{c}$.  
In conventional metals, even in thin homogeneous
films of conventional metals, $T_{\theta}$ is much greater than $T_c$,
which implies that phase fluctuations hardly effect the value of $T_c$.  

However, in the cuprate superconductors, $T_{\theta}$ is
comparable to $T_c$.   

\noindent{\bf A)} This conclusion can be reached on a priori 
theoretical grounds - in a doped antiferromagnet, we expect the 
effective bandwidth to be renormalized to be of order the 
antiferromagnetic exchange energy, $J$, and the density of pairs, 
$x/2$, to be determined by the density of doped holes.  The resulting estimate
$T_{\theta}\sim Jx/2$ is in the same ballpark as $T_{c}$ for 
 $J\approx 1500$K and $x\sim 0.1$ - 0.2.

\noindent{\bf B)}  More importantly, this conclussion is supported on empirical
grounds, if we use measured values of the London penetration depth to
deduce $T_{\theta}$.  Indeed, not only is $T_{\theta}$ comparable in
magnitude to $T_c$, it often exhibits similar trends with doping and
changes in other material properties, especially in the ``underdoped
region'' where $x$ is small;  this important correlation was originally
noticed by Uemura {\it et al}\cite{uemura}, and is called the ``Uemura relation.''
On the otherhand, in ``overdoped'' materials, $T_{\theta}$
overestimates $T_c$ by as much as a factor of three or four.  While this
is still a number of order 1, it suggests that ``optimal doping'' 
(the value of $x$ at which $T_{c}$ reaches a maximum value) may be 
loosely viewed as a crossover from an underdoped regime in which $T_{c}$ is
limitted by $T_{\theta}$ to an overdoped regime where it is
limitted by $T_{pair}$.

This same conclussion is reinforced by a variety of other empirical 
observations\cite{orensteinmillis}.  
In the first place, there is evidence that in optimally 
and underdoped cuprates, fluctuation effects are observed\cite{xy} in a rather 
broad range of temperatures ($\pm 10$\% or greater) about $T_{c}$
with a character and 
symmetry suggestive of the fluctuations in a classical XY model - this 
is the expected behavior of a system in which phase fluctuations 
dominate the physics in the neighborhood of $T_{c}$ and is quite 
different than the fluctuation effects seen in conventional metallic 
superconductors.  (Further evidence that superconducting fluctuations
of some sort
persist well above $T_{c}$ is presented in
Refs.  \ref{new} and \ref{ong}.)  At the same time, various measures of the zero 
temperature superconducting gap\cite{harris,fisher} suggest that  
$T_{pair}$ is a monotonically 
decreasing function of increasing $x$ with a magnitude large compared 
to $T_{c}$, especially in underdoped materials.  Moreover, the same sort of 
measurements invariably find that, for small enough $x$, the gap does 
not vanish at $T_{c}$, but rather evolves into a smeared and broadened 
``pseduo-gap.''  It is presently not unambiguous whether this large 
zero temperature gap and accompanying pseudo-gap can actually be 
interpretted as a superconducting (or pairing) gap\cite{ddw}.  However, if this 
interpretation is taken at face value, then the fact that $T_{pair}$ 
varies in the oposite sense as $T_{c}$ in the underdoped regime, and 
in the same sense in overdoped materials supports the identification 
of optimal doping as a crossover phenomenon.

\section{Theoretical Considerations}

The idea we pursue here is that we can raise the optimal $T_{c}$ in a 
two-component system by optimizing the pairing correlations in one 
subsystem and the superfluid stiffness in 
the other\cite{ekz1,ekz2,geballe}.   
We have previously carried out extensive theoretical studies of this 
idea in model one-dimensional systems, and shown that it 
works\cite{ekz1,ekz2}.  In particular, we demonstrated that strong 
superconducting correlations can be obtained by coupling a one 
dimensional electron gas  with a large 
spin-gap (a ``Luther-Emery liquid'') but small or 
vanishing superfluid density to a gapless one-dimensional electron gas 
(a ``Luttinger liquid'') with little or no spin-gap and a 
small superconducting susceptibility, 
but with a large Drude weight.  That is, it is possible to retain the
best aspects of each separate subsystem in the coupled system.

The theory of the one dimensional electron gas is well developped, 
and both weak and strong coupling limits are well understood 
theoretically, so these previous model studies are a sound basis for 
developping an intuition about real systems.  However, the 
theoretical analysis necessary to solve these models is somewhat 
technical, and aspects of the theory are special to one dimension.  
Here, we discuss the qualitative considerations that affect the ordering temperature of
such a two-component system in higher dimensions.  

Specifically, we will discuss a system
which consists  of alternating layers of "A" and "B" type.  In the absence of intelayer
coupling, the A-layers have a
large zero-temperature superconducting gap, $\Delta_A$, and
consequently a high mean-field transition temperature $T_{pair}\sim \Delta_A/2$, but 
a very small superfluid density so that their actual ordering temperature, $T_A \ll
\Delta_A/2$. Conversely, at $T=0$, the B-layers have a very small (or vanishing) gap, but
a large superfluid density (or, if not superconducting, a large Drude weight in the optical
conductivity).  We then ask the question:  What happens when we turn on moderately weak
interlayer couplings?

\subsection{The proximity effect and the amplitude of the order parameter}

One effect of single-particle tunnelling $t_{\perp}$ between planes is to produce the
usual  proximity effect.  What this means is that the electrons on plane A 
tend to leak onto plane B. This typically
has the combined effect of decreasing the pair-field magnitude on plane A and increasing
(or inducing)  pairing on plane B.   In the weak coupling limit, where $\Delta_A \ll
t_{\perp}$, this effect can be easily studied in the context of BCS mean-field theory.
Here, one first diaganolizes the single-particle Hamiltonian, including $t_{\perp}$, and
then evaluates the BCS gap equation using the exact single-particle eigenstates.  Because
the electron wave-functions are not localized on a single-plane, the amplitude for finding
two-particles near each other in plane A (where they experience a relatively strong
effective attraction) is reduced for those electrons that were originally localized on
plane A, and increased for those that were originally on plane B.  

If the Fermi surfaces
of planes A and B coincide, so that the exact eigenstates have equal amplitudes on each
plane, this effect is very significant, even for relatively small $t_{\perp}$ - roughly,
each plane develops a gap whose magnitude is proportional to $\Delta_A^2/W$, where $W$ is
the in-plane bandwidth.

If the Fermi surfaces of the two planes lie at substantially different locations in $\vec
k$ space, then the effect of small $t_{\perp}$ is small and can be computed in
perturbation theory.  Indeed, by explicit calculation it is possible to see that in this
limit,  $\Delta_A\to
\Delta_A[ 1 + {\cal O}(t_{\perp}/W)^2]$ and $\Delta_B\to\Delta_B + \Delta_A\times{\cal
O}(t_{\perp}/W)^2$.
Put another way, we are considering here the 
case in which single-particle tunnelling processes between plains are 
high energy, virtual processes which need not be considered 
explicitly.  Rather, the most important interactions for 
superconductivity are pair-tunnelling processes, in which a 
zero-momentum pair with energy near zero, tunnels between planes.  

\subsection{ Effects of pair-tunnelling on pairing}  

While for weakly correlated systems, correlated pair-tunnelling interactions
$J_{\perp}$ between layers are always weak, so the proximity effect dominates the
interlayer physics, in strongly interacting systems,  such pair-tunnelling interactions
are induced at low energies, and need not be weak compared to the single-particle effects.
Pair-tunnelling can affect the pairing scale (as discussed here) and can enhance
phase ordering, as discussed below.

Pair tunnelling  between plains can actually
lead to an  enhancement of the basic pairing scale.  The driving force for
 this effect is that, 
by pairing, the electrons can more easily delocalize between 
neighboring planes, and so lower their zero point kinetic energy.  
This effect can, by itself, be a principle mechanism of
pairing if, for either kinematic or dynamical reasons, 
single-particle tunnelling processes are greatly supressed while 
pair-tunelling remains strong.  This same physical idea underlies the interlayer 
tunnelling mechanism\cite{interlayer} of high temperature 
superconductivity,  and is also closely related to various ideas of multi-band
superconductivity\cite{multi,dhl}.  Its power as a pairing  mechanism in the one
dimensional context is well established -  it gives rise to what we\cite{ekz1,ekz2} 
have called the spin-gap proximity effect 
mechanism of pairing.  I think that some variant of this idea 
is likely to underlie the basic kinetic energy driven mechanism of 
pairing {\it within a plane}\cite{ekz1}.  

\subsection{ Effects of pair-tunnelling on phase ordering}

Josephson coupling between planar superconductors increases the stiffness of 
the system to phase twists, and so raises the phase ordering 
temperature.  To elucidate this effect, we consider the phase 
ordering in a set of coupled XY planes.  Clearly, the more strongly 
coupled the planes, the higher the phase ordering temperatures.  

For identical planes, this effect is more rapidly saturated 
than is commonly supposed.
We addressed this issue in a recent paper\cite{carlson} 
by studying the properties of layered 
classical XY models with nearest-neighbor coupling $J$ in a layer and 
$J_{\perp}$ between layers - in terms of measureable quantities at 
$T=0$:  $J=(\hbar c)^2/(16\pi e^2\lambda_{ab}^2)$ and 
$J_{\perp}/J\sim (\lambda_{ab}/\lambda_c)^{2}(\xi_{ab}/a_{c})^{2}$, 
where $\lambda_{ab}$ and $\lambda_{c}$  are the in-plane and interplane
components of the London penetration depth, $\xi_{ab}$ is the 
in-plane coherence length, and it is assumed that $a_{c}\ge \xi_{c}$.  
We solved these models using Monte-Carlo methods.  For $J_{\perp}=0$, 
$T_{c}=0.9J$;  it rises rapidly to $T_{c}=1.1 J$ for the relatively 
small value of $J_{\perp}=0.01J$.  (This is an impressive 20\% enhancement 
from a 1\% perturbation, and is the reason that two dimensional 
critical phenomena are so difficult to see in layered XY models.)
However, $T_{c}$ only rises to $T_{c}=1.3J$ for a further order of magnitude 
increase to $J_{\perp}=0.1J$.
Similarly, we modelled the case of multilayer materials by taking
$J_{\perp}=0.1J$ between the  $n$ layers
in a multiplet and $J_{\perp}=0.01J$ 
between multiplets;  here, for $n= 1,$ 2, 3, 4, $\infty$, we found that
$T_{c}= 1.09 J, 1.20 J, 1.24J,$ 1.26$J$, and $1.32 J$, respectively.  

While the general trend for $T_{c}$ to increase as identical layers 
are increasingly
coupled together is encouraging, it is clear from the above that this 
will not lead to very large $T_{c}$ enhancements.  To get large 
enhancements of the phase ordering temperature, it is necessary to 
increase the superfluid density per plane.  This can be
acheived in a two-component system, in which one of the components 
has a large phase stiffness.  For a broad range of 
interlayer couplings, a layered XY model with a large 
coupling, $J_{large}$, in one set, and a small coupling, $J_{small}$, 
in the other, will have a phase ordering transition at a critical 
temperature $T_{c}\sim J_{large}$.

There is one subtlety here, that bears mentioning.  At least in conventional BCS
superconductors, at temperatures near
$T_c$, the Josephson coupling between
two planes is proportional to the product of the order parameters on the two planes.
In expressing the physics of phase ordering in terms of an XY model with temperature
independent couplings, I have ignored all interference between the physics of pairing and
that of phase odering.  While for temperatures well less than $\Delta_A/2$ it may be
reasonable to neglect the temperature dependence of the superfluid stiffness in the A
planes, this approximation is certainly {\em not} valid for the interplane and
intra B-plane couplings at temperatures above the mean-field ordering
temperature of the B-planes.  To properly treat this full problem is beyond the scope of
any theory I know how to do.

\section{Materials Considerations}

Clearly, to make a two component system superconduct at high 
temperatures, it is better to use as constituents as good
superconductors as possible.  The new idea, here, is that a
higher transition temperature than is possible with either
constitutent alone can be obtained by acheiving a particularly
high pairing scale in one constitutent and a sufficiently large
phase stiffness in the other.  This is a very general strategy,
and can be applied to the design of many materials.  There are a few
obvious, but important aspects of this stategy, indicated in the theoretical discussions
above, which bear summarizing:  

Firstly,
in choosing the material from which the  pairing scale derives,
the first goal is to acheive a high pairing scale (superconducting
gap), but for given gap magnitude, the higher the superfluid density,
the better.  Conversely, the material with the high  superfluid
density will be more effective if it has a substantial pair-field
susceptibility at elevated temperatures.  In short, we may not want
to go to extremes.

Secondly, in coupling the neighboring planes, we want to supress
low energy single-particle tunnelling (which will tend to decrease the 
maximum pairing scale) but not at the expense of pair-tunnelling, 
which is necessary to couple the phase fluctuations in the two 
components.  This can be the result of kinematics ({\it i.e.} a 
mismatch in Fermi momenta in the two sytems\cite{ekz1,ekz2}) or
dynamics\cite{interlayer,vadim}.  (See, aslo, \ref{sudip}.)  

\subsection{A Model System}

An interesting study has been carried out\cite{merchant} on a model 
system by Merchant {\it et al}, in which the crossover from phase 
ordering to pairing can be observed directly.  

The starting material 
for this study is a film of granular lead with sufficiently small lead 
coverage that, resistively, it is an insulator at low temperatures.  
However, transverse tunnelling into this starting film reveals the 
presence, locally, of a well developped superconducting gap below the 
bulk $T_{c}$ of lead;  the insulating behavior is unambiguously the 
result of quantum and thermal phase fluctuations in this film, and is 
reflective of the fact that the superfluid density (in this case, 
determined by the typical Josephson coupling between grains) is 
small.  

From this starting point, a sequence of films were made by adding 
increasing amounts of silver.  Silver is thought to be depositted as 
a homogeneous covering film, even at very low coverages.  Two effects 
of increasing silver coverage are observed.  Firstly, the silver 
enhances the superfluid stiffness, resulting, at small coverage, in a 
resistive $T_{c}$ (defined as the point at which the resistance drops 
to a suitably small fraction of its normal state value) which is an 
increasing function of the amount of silver depositted.  However, the 
proximity effect results in a monotonic decrease of the 
superconducting gap observed in tunelling, which is slow at first, 
but then more rapid with increasing silver coverage.  Eventually, 
this results in a reversal of the trend observed in the resistivity, so that
$T_{c}$ is found to reach a maximum (at an ``optimum silver 
coverage'') and then drop rapidly with increasing silver coverage.

There are many subtleties of these experiments that are still not well 
understood.  For instance, although in the ``overdoped'' regime, the 
superconducting transition is sharp and the superconducting state 
itself well defined, on the underdoped side the transition is 
rounded and it appears, for a range of silver coverage, that a zero 
resistance state is never acheived, even in the limit $T\to 0$.  
However, the basic trends observed in these experiments
serve as a proof in a real material that the basic 
strategy for $T_{c}$ enhancement proposed here can work!

\subsection{Considerations specific to the cuprates}

In a crystal with one layer per unit cell, unless the crystalline 
symmetry is spontaneously broken, each layer will have the same doped 
hole concentration.  This remains true in a bilayer material which possesses a 
reflection plane or screw symmetry which links the planes.  This 
situation applies, to the best of my knowledge, to all the bilayer 
cuprate superconductors that have been studied to date.  One 
could imagine, however, bilayer materials of lower symmetry
designed so that the doped 
hole concentration in each layer can be varied separately.  All trilayer 
and four-layer materials have two crystallagraphically inequivalent 
copper-oxide planes, which should therefore be expected, a priori, to 
have different doped hole concentrations\cite{stasio}.  Thus, there already exist 
materials that realize, to some extent, the scenario envisaged here.  

This simple observation leads to the suggestion that 
the enhanced $T_{c}$'s observed in 
three and four layer materials, for which various other mechanisms 
have already been suggested\cite{other}, may be explained by the present ideas.
To test this idea, one should examine, with various local probes, the 
differences in the doped hole concentrations in the different layers.  
One striking observation is that in some of these materials, notably 
in the three and four layer materials\cite{jover,tallon}, {\HgCun} and {\TlCun} with $n=3$
and 4,
$T_{c}$ is remarkably pressure dependent.  A potentially  testable prediction based on the
present analysis is that this  dependence is a consequence of a pressure dependent charge
transfer  between planes\cite{related}.  If so, it should lead to a charge distribution
which  looks more and more optimal (in the sense described below) as $T_{c}$ 
rises.

What then is the optimal distribution of doped holes between two 
coupled layers.  From the observation that the superfluid density in 
overdoped cuprates does not increase with increasing hole 
concentration, although the gap size decreases markedly, we deduce 
that the ``high superfluid density layer'' should be optimally, or at 
most very slightly overdoped.  On the otherhand, the gap scale appears 
to rise rapidly with underdoping, at least according to some 
measures.  Thus, the  ``large pairing scale layer'' should ideally 
be substantially underdoped.  However, the empirical trends are less 
clear with extreme underdoping.  For one thing, in this range, 
screening becomes very poor and the effects of disorder appear to be 
much enhanced.  Thus, at least at first, extreme underdoping should 
probably be avoided as well.

\noindent{This work is an outgrowth of collaborative research with
many coleagues, but especially with V. J. Emery, O. Zachar, E. W. Carlson, D. Orgad,
and E. Fradkin.}  


\end{multicols}

\end{document}